\documentclass[]{pasj01}
\draft

\Received{}
\Accepted{}
 
\usepackage{lineno}
 
\begin{document} 

\title{Absolute Properties of the Oscillating Eclipsing Algol XZ Ursae Majoris }

\author{Jae Woo \textsc{Lee}\altaffilmark{1,2}%
}
\altaffiltext{1}{Korea Astronomy and Space Science Institute, Daejeon 34055, Republic of Korea}
\altaffiltext{2}{Astronomical Institute, Faculty of Mathematics and Physics, Charles University in Prague, 180 00 Praha 8, V Hole\v sovi\v ck\'ach 2, Czech Republic}
\altaffiltext{3}{Institute for Astrophysics, Chungbuk National University, Cheongju 28644, Republic of Korea}
\altaffiltext{4}{Department of Astronomy and Space Science, Chungbuk National University, Cheongju 28644, Republic of Korea}
\altaffiltext{5}{National Astronomical Research Institute of Thailand, Chiang Mai 50200, Thailand}
\email{jwlee@kasi.re.kr}

\author{Kyeongsoo \textsc{Hong}\altaffilmark{1}}
\author{Hye-Young \textsc{Kim}\altaffilmark{3,4}}
\author{Marek \textsc{Wolf}\altaffilmark{2}}
\author{Jang-Ho \textsc{Park}\altaffilmark{1}}
\author{Pakakaew \textsc{Rittipruk}\altaffilmark{5}}

{}

\maketitle

\begin{abstract}
It is known from archival TESS data that the semi-detached Algol system XZ UMa is one of the candidate binary stars exhibiting 
short-period oscillations. We secured new high-resolution spectroscopic observations for the program target to better 
understand its binary and pulsation properties. From the echelle spectra, the radial velocities (RVs) of the eclipsing pair 
were derived, and the atmosphere parameters of the primary component were measured to be $v_{\rm A}\sin$$i = 80\pm7$ km s$^{-1}$, 
$T_{\rm eff,A}$ = $7940\pm120$ K, and [M/H] = $-0.15\pm0.20$. The combined solution of our double-lined RVs and the TESS data 
provides robust physical parameters for XZ UMa with mass and radius measurement precision of better than 2 \%. 
The outside-eclipse residuals from a mean light curve in the 0.002 phase bin were used for multifrequency analyses, and 
we extracted 32 significant frequencies (22 in $<$ 5.0 day$^{-1}$ and 10 in 39$-$52 day$^{-1}$). The low frequencies may be mostly 
aliasing sidelobes, while six of the high frequencies may be pulsation signals arising from the detached primary located inside 
the $\delta$ Sct domain. Their periods, pulsation constants, and pulsational-orbital-period ratios indicate that the mass-accretion 
primary star is a $\delta$ Sct pulsator and, hence, XZ UMa is an oscillating eclipsing Algol. 
\end{abstract}


\section{Introduction}

In our Galaxy, there are more than twice as many binary or multiple stars as single stars (Duquennoy \& Mayor 1991). 
Understanding such systems is almost the same as understanding stellar physics. Binary stars still serve as the primary object 
for obtaining fundamental data. In particular, spectroscopic and eclipsing binaries (SEBs) allow for direct and independent 
determination of the absolute dimensions of each star and the distances to these systems, without any assumptions. 
This is accomplished using a combined analysis of two types of observations: radial velocities (RVs) and light curves 
(Hilditch 2001; Torres et al. 2010). The masses and radii of well-studied SEBs can be measured with a precision of better than 
2 \%, and the results employed to test and improve stellar evolution models (Southworth 2015; Serenelli et al. 2021). 
Recently, a considerable number of eclipsing binaries (EBs) have been observed and newly discovered in the time-series 
photometric surveys of Kepler (Kirk et al. 2016) and TESS (Pr\v sa et al. 2022). The impact of these space missions on EBs was 
detailed in Southworth (2021), who showed that various types of pulsating variables were found at almost all stages of EB evolution 
through high-precision data unachievable from the ground. 

Asteroseismic modeling of pulsating stars makes it possible to probe their interiors, from cores to envelopes, providing rigorous 
constraints on stellar theory. Double-lined SEBs which contain a pulsating component provide valuable information that improves 
our understanding of stellar physics, based on the binary and pulsation properties. A recent study by Gaulme \& Guzik (2019) reported 
that about 10 \% of almost 3000 Kepler EBs display evidence of pulsations in their ultra-precise light curves. The stellar pulsations 
in close binaries may be affected by the mass accretion and tidal force between their components (Mkrtichian et al. 2018; 
Bowman et al. 2019; Guo 2021). One typical example is the oEA (oscillating eclipsing Algol) class: the mass-gaining primary components 
in semi-detached Algols exhibit $\delta$ Sct-like pulsations through mass accretion from their lobe-filling companions 
(Mkrtichian et al. 2004). In general, $\delta$ Sct variables are pulsating A/F stars located where the Cepheid instability strip 
intersects the main sequence, and they oscillate in pressure ($p$) modes with periods of 0.02$-$0.3 days and brightness variations 
of $<$ 0.1 mag (Breger 1979, 2000). 

The pulsating SEBs are ideal targets for asteroseismic study, and their numbers have been significantly increased by space-based 
survey observations (Gaulme \& Guzik 2019; Kahraman Ali\c cavu\c s et al. 2022; Shi et al. 2022). Nonetheless, it has not been 
possible to measure most of them with reliable fundamental stellar parameters, because of the absence of spectroscopic observations. 
High-resolution time-series spectra are needed to study the interesting variables in detail. We have been conducting 
spectroscopic observations of pulsating SEBs using 2-m class telescopes in Korea and Thailand (Hong et al. 2015, 2019). Based on 
archival TESS data, we include the pulsating SEB candidate XZ Ursae Majoris (TIC 318217844; Gaia DR3 1017991469167531136; 
TYC 3429-1530-1; $T_{\rm p}$ = $+$9.996; $V\rm_T$ = $+$10.507, $(B-V)\rm_T$ = $+$0.218; hereafter XZ UMa) in our observation list. 

XZ UMa was listed as a semi-detached Algol (EA/SD) with a spectral type of A5+F9 and an orbital period of 1.22232 days in the 4th GCVS 
(Kholopov et al. 1992). Kim et al. (2003) did a photometric survey to check for pulsations in the system, but found no evidence of them. 
The first comprehensive study of this target was carried out by Nelson et al. (2006). The authors obtained the light curves in 
the $BVI_{\rm c}$ band and the double-lined RVs from 11 spectra, and estimated the primary star's temperature to be $7766\pm240$ K 
using the color index ($B-V$) = +0.20$\pm$0.04 from the Tycho Catalogue (ESA 1997). The binary model parameters and absolute dimensions 
of XZ UMa were determined from a simultaneous modeling of their own observations, indicating that the target star is a semi-detached Algol. 
In the modeling, they found no significant values of both non-synchronous rotation and third light. Since then, Soydugan et al. (2011) 
showed that the orbital period of XZ UMa has changed by a combination of a downward-opening parabola and a sinusoid, and proposed that 
the periodic variation is produced by the possible existence of a tertiary component with a period of 29.4 years and a minimum mass of 
0.51 $M_\odot$ in an elliptical orbit with eccentricity 0.56. In addition, these authors detected a third light of about 5 \% by means of 
re-analyzing the $V$ light curve of Nelson et al. (2006), which could be evidence for the circumbinary object deduced in their period study. 

This article is the seventh in a series of papers on the oEA stars (Hong et al. 2015, 2017, 2019; Koo et al. 2016; Park et al. 2020; 
Lee et al. 2023). We present the archival TESS photometry (Section 2) and our high-resolution spectroscopy (Section 3) of XZ UMa, 
measure the fundamental parameters of each component (Section 4), and show that the mass-accreting primary is a $\delta$ Sct-type 
pulsating star via multifrequency analyses (Section 5). Throughout the paper, we express each measurement's error as a value of 
1$\sigma$ (standard deviation) unless it is noted.

\section{TESS Photometry and Eclipse Times}

The ultra-precise TESS observations of XZ UMa were secured during Sector 21 from 2020 January 21 to February 18 (BJD 2,458,870.43 
$-$ 2,458,897.79). In this study, we adopted the 2-min cadence SAP data taken from the TESS EB catalogue\footnote{http://tessEBs.villanova.edu} 
and converted the normalized fluxes to relative magnitudes. A detailed description of this catalogue was given in Pr\v sa et al. (2022). 
The light curve of the target star is illustrated in Figure 1 as magnitude unit versus orbital phase, and it consists of 18,631 points 
with the black circles. The CROWDSAP\footnote{target-to-total flux ratio within the photometric aperture} value of the TESS data is 
0.99909002, so other stars would not have affected the flux measurements for XZ UMa. The depth difference between the primary and 
secondary eclipses indicates there is a considerable temperature difference between both components, typical of classical Algols. 
Moreover, the space-based observations display the O'Connell effect whereby Max I at 0.25 phase is $\sim$0.006 mag brighter than Max II 
at 0.75 phase. The light asymmetry may originate from magnetic starspot activity on the cool component, whose envelope is considered 
a convective atmosphere. 

A total of 43 XZ UMa eclipses were recorded from the time-series TESS photometry of about 27 days, and their times of minima were 
derived with the method of Kwee \& van Woerden (1956). The results are presented in Table 1, where Min I and Min II represent 
the primary and secondary times of minima, respectively. By means of a least-squares fit applied to the TESS eclipse timings, 
we obtained the following linear ephemeris: 
\begin{equation}
\mbox{Min I} = \mbox{BJD}~ 2,458,882.843704(5) + 1.2223025(8)E.
\end{equation}
The 1$\sigma$-errors for each term are presented in the parentheses. The binary star's period corresponds to an orbital frequency 
of $f_{\rm orb}$ = 0.8181281 $\pm$ 0.0000005 day$^{-1}$.

Using all available mid-eclipse times, we show the current $O-C$ diagram of XZ UMa in the top panel of Figure 2. In this article, 
we collected a total of 396 minimum epochs. Most of them can be found in the $O-C$ Gateway database\footnote{http://var2.astro.cz/ocgate/index.php?lang=en}, 
Information Bulletin on Variable Stars (IBVS), and Open European Journal on Variable Stars (OEJV). Applying the quadratic fit to 
all the minima, we calculated the following ephemeris: 
\begin{equation}
 C = \mbox{HJD}~ 2,440,291.5666(9) + 1.22230995(3)E - 2.40(5) \times 10^{-10}E^2. 
\end{equation}
The result is presented as a red parabola in Figure 2. The $O$--$C_{\rm quad}$ values from the quadratic ephemeris are shown in 
the other panels of this figure. The downward parabola indicates a decay of the orbital period at the rate of $-$1.43$\times$10$^{-7}$ day yr$^{-1}$. 

The timing residuals obtained since 2000 for more precise photoelectric or CCD minima are presented in the middle panel of Figure 2. 
Our TESS timings are marked as a bulk of points near epoch 15,100. One can see a semi-regular oscillation with 
a characteristic length of about 10 years and an amplitude up to 0.004 days. Because the $O-C$ values are not strictly periodic, 
it is difficult to simply explain this variation as a light-time effect (Irwin 1952) caused by a circumbinary companion. Moreover, 
these data still have relatively large scatter. Using only the precise TESS minima, the detailed $O-C$ diagram of XZ UMa is shown on 
the bottom panel of Figure 2. This diagram covers exactly 22 epochs obtained in Sector 21 of the TESS observations. 
A systematical shift of all secondary eclipses of about +0.001 days is caused very probably by the existence of a dark spot persisting 
on the less massive secondary component, which will be discussed in Section 5. 

XZ UMa belongs to the group of well-known short-period Algols showing a quasi-sinusoid superposed on a long-term period decrease 
(Lee et al. 2023). Neither of these two variations have yet been clearly explained.

\section{New Spectroscopy and Spectral Analysis}

Because there is a large difference in luminosity between the XZ UMa components, it is not easy to obtain its double-lined RV curves. 
High-resolution spectra with sufficient signal-to-noise (S/N) ratios are required for the RV measurements of the faint companion. 
The spectroscopic observations of XZ UMa were carried out on eight nights between 2021 February and December using the BOES 
(Bohyunsan Observatory Echelle Spectrograph; Kim et al. 2007) and the 1.8-m telescope at Bohyunsan Optical Astronomy Observatory (BOAO) 
in Korea. The spectrometer recorded the high-resolution spectra in the range of 3600$-$10,200 $\rm \AA$ across 75 spectral orders. 
The resolving power and exposure time of the XZ UMa spectra were given as 30,000 and 30 min, respectively. 
Their S/N ratios were extracted between 4000 $\rm \AA$ and 5000 $\rm \AA$, and they were typically placed at 35$-$40. A total of 
54 observed spectra were acquired for the XZ UMa system. The data reductions were performed using the interactive package IRAF, which 
included bias and flat corrections, wavelength calibration, extraction, and normalization to the continuum (Hong et al. 2015). 

Trailed spectra are considered a useful tool and show the positional changes of absorption lines due to binary orbital motion 
(Hong et al. 2015, 2019). The phase-folded trailed spectra for XZ UMa were constructed to find measurable absorption lines for 
its double-lined RVs. We examined them over the entire wavelength by visual inspection, and found the obvious S-wave features of 
the binary components A and B around Fe I $\lambda$4957. Figure 3 shows the trailed spectra in this region (upper panel) for 
XZ UMa, and a sample spectrum observed at phase 0.256 (lower panel). The Fe I absorption lines of all individual spectra were 
fitted using double Gaussian profiles with the deblending tool within IRAF/splot. The resulting RV measurements of XZ UMa are listed 
in Table 2 and are plotted in Figure 4, together with those of Nelson et al. (2006). In this figure, it can be seen that our data 
uniformly covered most orbital phases. 

To measure the atmosphere parameters of the binary star, it is necessary to obtain higher S/N spectra for each component than 
we observed. To do so, we extracted the disentangled spectrum of each component from the observed spectra of XZ UMa, blended by 
the orbital motion, applying the FDB\textsc{inary} code\footnote{http://sail.zpf.fer.hr/fdbinary} of Iliji\'c et al. (2004). 
Figure 5 displays the disentangled spectrum of the primary component for the seven spectral regions (H$_{\rm \delta}$, 
Ca I $\lambda$4226, H$_{\rm \gamma}$, Fe I $\lambda$4383, Mg II $\lambda$4481, H$_{\rm \beta}$, Fe I $\lambda$4957) used in 
this paper, where their S/N ratios were about 60$-$80. In contrast, we did not obtain the reconstructed spectrum of the faint companion 
with sufficient S/N ratios to estimate its reliable atmospheric parameters. 

The projected rotational velocity $v_{\rm A}$sin$i$, effective temperature $T_{\rm eff,A}$, and metallicity [M/H] of the primary star 
were determined by the means of fitting its disentangled spectrum with the use of the iSpec code (Blanco-Cuaresma et al. 2014a). 
In the run, the synthetic spectra were computed by employing the Kurucz stellar models (Castelli \& Kurucz 2003) and the SPECTRUM code 
(Gray \& Corbally 1994). The list of atomic lines was provided by the third version of the VALD (Vienna Atomic Line Database; 
Ryabchikova et al. 2015). The micro and macro turbulence velocities were calculated by empirical relations implemented in the iSpec code, 
which are based on the FGK benchmark stars from the Gaia-ESO Survey (Blanco-Cuaresma et al. 2014b; Jofr\'e et al. 2014). 
Initial temperature and surface gravity in the code were set to be $T_{\rm eff,A} = 7766\pm240$ K (Nelson et al. 2006) and 
$\log g_{\rm A}$ = 4.3 (cf. Section 4), respectively. 

First of all, the $v_{\rm A}$sin$i$ value was derived by fitting the Mg II $\lambda$4481 line profile with the synthetic spectra, 
where the Mg II absorption is the most reliable photosphere line for determining stellar rotation (e.g. Royer et al. 2002a,b). 
Then, we calculated the two atmosphere parameters of $T_{\rm eff,A}$ and [M/H] by fixing the rotation parameter obtained in 
the previous step and simultaneously analyzing the remaining six absorption lines useful for the temperature classification of 
A-type stars (Gray \& Corbally 2009). This process was repeated iteratively until the best fit was obtained. The best-fitting model 
has the following parameters: $v_{\rm A}\sin$$i = 80\pm7$ km s$^{-1}$, $T_{\rm eff,A}$ = $7940\pm120$ K, and [M/H] = $-0.15\pm0.20$.  
The error estimates within iSpec were probably underestimated, and they were adopted from the standard deviation of the optimal 
values calculated from each spectral region. The spectroscopic temperature was within the margin of error with the $7766\pm240$ K 
value obtained photometrically from the color index of ($B-V$) = +0.20$\pm$0.04 (Nelson et al. 2006).

\section{Binary Modeling and Absolute Dimensions}

A photometric-spectroscopic study by Nelson et al. (2006) pointed out that XZ UMa is a semi-detached classical Algol system. 
Thus, for the binary modeling, we employed all of our double-lined RV and TESS light curves with mode 5 of the Wilson-Devinney (W-D) 
synthesis code (Wilson \& Devinney 1971; Van Hamme \& Wilson 2007). The binary code reasonably represents the geometric distortions 
of the component stars based on the Roche model, and determines the orbital and stellar parameters that best fit between observations 
and models. Our approach to analyzing the two datasets is almost identical to that for oscillating EBs OO Dra (Lee et al. 2018) and 
V404 Lyr (Lee et al. 2020a). Through an iterative method, the archival TESS data were analyzed with our double-lined RV measures. 

In the modeling process, the mean surface temperature $T_{\rm A}$ of the XZ UMa primary was held fixed at $7940\pm120$ K as yielded 
by our disentangled spectrum analysis. The bolometric albedo $A_{\rm A}$ and gravity-darkening exponents $g_{\rm A}$ for XZ UMa A were 
both given as 1.0, and a synchronous rotation for XZ UMa B was assumed, $F_{\rm B}$ = 1.0. In contrast, the rotation-to-orbit velocity 
ratio for XZ UMa A was taken as $F_{\rm A}$ = 1.10$\pm$0.10, as the observed $v_{\rm A}\sin$$i$ was faster than the synchronized velocity 
$v_{\rm A,sync}$ = $73.0\pm1.4$ km s$^{-1}$ calculated from $2 \pi R_{\rm A}$/$P_{\rm orb}$. The limb-darkening (LD) parameters $x$ and 
$y$ with the logarithmic law were initialized from those tabulated and updated by Van Hamme (1993). The parameters adjusted in this section 
were: linear ephemeris $T_0$ and $P_{\rm orb}$, semi-major axis $a$, system velocity $\gamma$, mass ratio $q$, inclination angle $i$, 
secondary's temperature $T_{\rm B}$, albedo $A_{\rm B}$, and gravity-darkening exponents $g_{\rm B}$, primary's potential $\Omega _{\rm A}$ 
and luminosity $L_{\rm A}$, third light $l_3$, and linear LD term $x_{\rm A,B}$. Model starspots are applied to fit the light asymmetry 
between the Max I and Max II quadratures. 

The TESS light and our RV curves were modeled simultaneously until each parameter's correction was smaller than its corresponding 
1$\sigma$ value. The quality of each synthesis was checked by computing the sum of the weights of the squared residuals between 
observations and models. As a final result, the optimal binary star parameters are summarized in Table 3, and the model curves from 
them are displayed as red solid lines in the top panel of Figures 1 and 4. 
The Roche-lobe geometrical representations of XZ UMa are given in Figure 6. Obtained parameters during the binary modeling confirm 
that the eclipsing pair of XZ UMa has a circular-orbit semi-detached configuration, and show that the detached primary fills 
its inner critical lobe in about 65 \%. As described in Section 2, the CROWDSAP factor of 0.99909002 indicates that XZ UMa was not 
observed in the same pixel with other stars. Hence, the possible source of $l_3$ = 0.054 could be a circumbinary object, as proposed 
by Soydugan et al. (2011). The light and velocity residuals corresponding to our binary model are drawn at the middle and bottom of 
Figures 1 and 3, respectively. As one can see here, the model fits the observed RV curves satisfactorily, while the TESS residuals show 
visible trends. There may be many possible explanations for this mismatch between observations and models, such as model assumptions and 
irradiation mechanism used in the W-D code, heat redistribution in the component atmospheres, magnetic activities in the cool secondary, 
gas streams and/or circumbinary dust cloud. At present, it is difficult to clearly state which of these possibilities is the main cause. 
Nonetheless, we speculate that the light discrepancy may be attributed to binary interactions which are currently unknown or not included 
in the W-D model (see Southworth \& Bowman 2022). 

The absolute dimensions for XZ UMa were obtained from the comprehensive and detailed analysis of both our spectroscopic and TESS 
photometric data. For this, the surface temperature of the sun was set to be $T_{\rm eff}$$_\odot$ = 5780 K, and its bolometric magnitude 
to be $M_{\rm bol}$$_\odot$ = +4.73. To obtain absolute visual magnitudes $M_{\rm V}$, bolometric corrections BCs were adopted from 
the empirical equation using the effective temperature presented in Table 1 of Torres (2010). The calculation results are illustrated in 
Table 4 for comparison with those of Nelson et al. (2006). Our radii and surface gravities are consistent with previous results in 
the margins of their errors, while the present masses for the primary and secondary stars are both about 13 \% heavier. 

The distance $d$ to XZ UMa was calculated from the standard distance-modulus equation $V-A_{\rm V}-M_{\rm V}=5\log{d}-5$. Here, 
$V$ and $A_{\rm V}$ $\simeq$ 3.1$E(B-V)$ are the apparent magnitude and interstellar extinction of the EB system, respectively. 
We took $V$ = 10.27 $\pm$ 0.05 and $E$($B-V$) = 0.014 $\pm$ 0.004 from the TESS Input Catalogue revised by Stassun et al. (2019). 
Because XZ UMa is a potential triple system (Soydugan et al. 2011) and our analysis detected $l_3$ = 0.054 in the TESS band, 
the $V$ magnitude may include the third light source. Thus, we used $V_{\rm bin}$ = 10.32 for the binary star applying 
$l_{\rm 3,V}$ = 0.048 obtained in the $V$ band by Soydugan et al. (2011). The EB distance of 489 $\pm$ 20 pc presented in this article 
corresponds well to the Gaia DR3 distance of 483 $\pm$ 4 pc inverted from its parallax 2.070$\pm$0.017 mas (Gaia Collaboration 2022). 
For population membership, the Galactic kinematics of XZ UMa were computed to be $U = 34.7 \pm 0.5$ km s$^{-1}$, 
$V = 230.1 \pm 0.1$ km s$^{-1}$, and $W = 2.1 \pm 0.5$ km s$^{-1}$, following the procedure used by Lee et al. (2020b). The location 
of the EB system in the $U-V$ plane falls amid the thin-disk stars classified by Pauli et al. (2006). This means that XZ UMa has 
the kinematics of the thin-disk population.

\section{Pulsational Characteristics}

From its position on the Hertzsprung-Russell diagram, the XZ UMa primary is a candidate $\delta$ Sct-type pulsator in EBs. 
Because XZ UMa A and B are obscured by each other during the eclipses, we employed the out-of-eclipse light curve residuals 
between phases 0.10$-$0.40 and 0.60$-$0.90 for periodogram analysis. First of all, the residuals from the W-D model were applied 
to the PERIOD04 (Lenz \& Breger 2005) to show a frequency spectrum in the top panel in Figure 7, where many orbital harmonics 
are clearly visible. They may come mainly from an imperfect binary star model.  
To better detect the presence of pulsations in XZ UMa, we used a mean light curve, instead of the W-D fit to the TESS data 
(cf. Lee et al. 2023). For this, we examined the mean light curves and their residuals formed in various phase intervals of 
0.03, 0.02, 0.01, 0.005, 0.002, 0.001, 0.0007, 0.0005, and 0.0001, and the fifth of them was chosen. The residual TESS data 
after subtracting the chosen mean curve are displayed in the bottom panel of Figure 1. It showed that the binary signals were 
well subtracted and the residual dispersion has decreased considerably.

The light residuals from the mean curve in the 0.002 phase bin were analyzed according to iterative and simultaneous pre-whitening 
using the PERIOD04 program, as in Lee et al. (2014). As a consequence of the frequency analysis, we revealed a total of 32 
significant signals, which are summarized in Table 5. The amplitude spectra of XZ UMa for the mean-curve residuals are plotted in 
the second to bottom panels of Figure 7. As shown in the table and figure, all of the frequency signals fall into two domains: 
22 in $<$ 5.0 day$^{-1}$ and 10 in 39$-$52 day$^{-1}$. We used the Rayleigh criterion of 1/$\Delta T$ = 0.04 day$^{-1}$, to check 
whether the extracted frequencies are possible orbital harmonics and combinations. This consequence is shown in the remark column 
of Table 5. The $f_{4}$ frequency corresponds to the time base ($\Delta T$ =27.35 days) for the TESS data used, and $f_{7}$ and 
$f_{11}$ seem to be related to the rotational frequency of $f_{\rm rot}$ = 0.899 day$^{-1}$. We suspect that almost all of the signals 
in the low-frequency domain may be alias effects attributed to only using the outside-eclipse residuals and the imperfect removal of 
systematic data trends. 

On the other hand, most of the high frequencies around 48 day$^{-1}$ may result from the primary component that resides inside 
the $\delta$ Sct domain. Except for four possible combinations ($f_{26}$, $f_{27}$, $f_{30}$, $f_{31}$), the remaining six high 
frequencies could be identified as pulsating signals. We applied these frequencies $f_i$ and the primary's density $\rho_{\rm A}$ to 
the well-known equation of $Q_i$ = $f_i$$\sqrt{\rho_{\rm A} / \rho_\odot}$, and computed the pulsation constants of $Q_6$ = 0.013 days, 
$Q_{12}$ = 0.013 days, $Q_{18}$ = 0.014 days, $Q_{23}$ = 0.013 days, $Q_{24}$ = 0.014 days, and $Q_{29}$ = 0.016 days. 
The frequency signals could be classified as fourth or fifth overtone modes of $\delta$ Sct stars, respectively (Fitch 1981). 
These $Q$ values and the pulsation periods of $P_{\rm pul}$ = 0.020$\sim$0.026 days imply that the primary star is oscillating in 
the pressure ($p$) modes of typical $\delta$ Sct variables (Breger 2000).

\section{Discussion and Conclusions}

For the pulsating SEB XZ UMa, we obtained time-series echelle spectra using the 1.8-m reflector and BOES spectrograph at BOAO. 
The spectroscopic observations were combined and analyzed in detail with the highly precise photometric data provided by 
the TESS space mission. We measured the double-lined RVs of XZ UMa by fitting the Fe I $\lambda$4957 absorption lines 
of the BOES spectra with two Gaussian functions. Using the iSpec code, the rotational velocity, surface temperature, and 
metallicity of the primary component were found to be $v_{\rm A}\sin$$i = 80\pm7$ km s$^{-1}$, $T_{\rm eff,A}$ = $7940\pm120$ K, 
and [M/H] = $-0.15\pm0.20$, respectively. Our observed $v_{\rm A}\sin$$i \approx v_{\rm A}$ indicates that the XZ UMa A primary 
is likely currently in a super-synchronous state. 

The combined analysis of the spectroscopic and photometric measurements indicates that our target is a semi-detached Algol 
with physical properties of $M_{\rm A} = 2.177 \pm 0.040$ $M_\odot$, $R_{\rm A} = 1.764 \pm 0.034$ $R_\odot$, and $L_{\rm A} = 
11.08 \pm 0.79$ $L_\odot$ for XZ UMa A and $M_{\rm B} = 1.356 \pm 0.026$ $M_\odot$, $R_{\rm B} = 2.481 \pm 0.045$ $R_\odot$, 
$T_{\rm eff,B}$ = $5162 \pm 53$ K, and $L_{\rm B} = 3.92 \pm 0.21$ $L_\odot$ for XZ UMa B, respectively. The masses and radii 
were yielded with measurement precisions better than 2 \%. The detached primary fills its critical lobe by 65 \% and is 
0.72 $R_\odot$ smaller than the lobe-filling companion. Our fundamental parameters were determined from a sufficient number of 
high-resolution spectra with much greater observational weight than that of Nelson et al. (2006). Thus, we prefer our results 
presented in this article. The kinematic data of XZ UMa indicate that the binary system belongs to the thin-disk population in 
the $U-V$ diagram, which matches our metallicity measurement in a disk population. 

For multifrequency analyses, we made a mean curve in the 0.002 phase bin and analyzed the residual lights after subtracting it 
from the TESS data. A total of 32 significant frequencies were uncovered, of which the low frequencies below 5.0 day$^{-1}$ can 
be considered aliasing sidelobes due to incomplete removal of the binarity and systematic trends in the observed TESS data. 
In contrast, the high frequencies between 39 and 52 day$^{-1}$ were located within the realm of $\delta$ Sct pulsations, and 
many of them may be oscillating signals excited in the XZ UMa A primary. The six signals of $f_6$, $f_{12}$, $f_{18}$, $f_{23}$, 
$f_{24}$, and $f_{29}$ seem to be possible independent pulsation frequencies. Their periods and pulsation constants are in ranges 
of $28 < P_{\rm pul} < 37$ min and $0.013 \le Q \le 0.016$ days, respectively. The pulsational-orbital-period ratios for 
the high frequencies are $P_{\rm pul}/P_{\rm orb}$ = 0.016$\sim$0.021 within the 0.09 limit for the $\delta$ Sct-type stars in 
eclipsing systems (Zhang et al. 2013). All of these indicate that XZ UMa is a $\delta$ Sct EB in a semi-detached configuration 
and hence the mass-accreting primary component is a new example of an oEA star.

\begin{ack}
We would like to thank the BOAO staffs for assistance during our spectroscopy. This paper includes data collected by the TESS mission. 
Funding for the TESS mission is provided by the NASA Explorer Program. This research has made use of the Simbad database maintained 
at CDS, Strasbourg, France, and was supported by the KASI grant 2023-1-832-03 and the project Cooperatio - Physics of the Charles 
University in Prague. H.-Y.K. was supported by the grant number 2019R1A2C2085965 from the National Research Foundation (NRF) of Korea.
\end{ack}

\clearpage
\begin{figure}
\begin{center}
\includegraphics[]{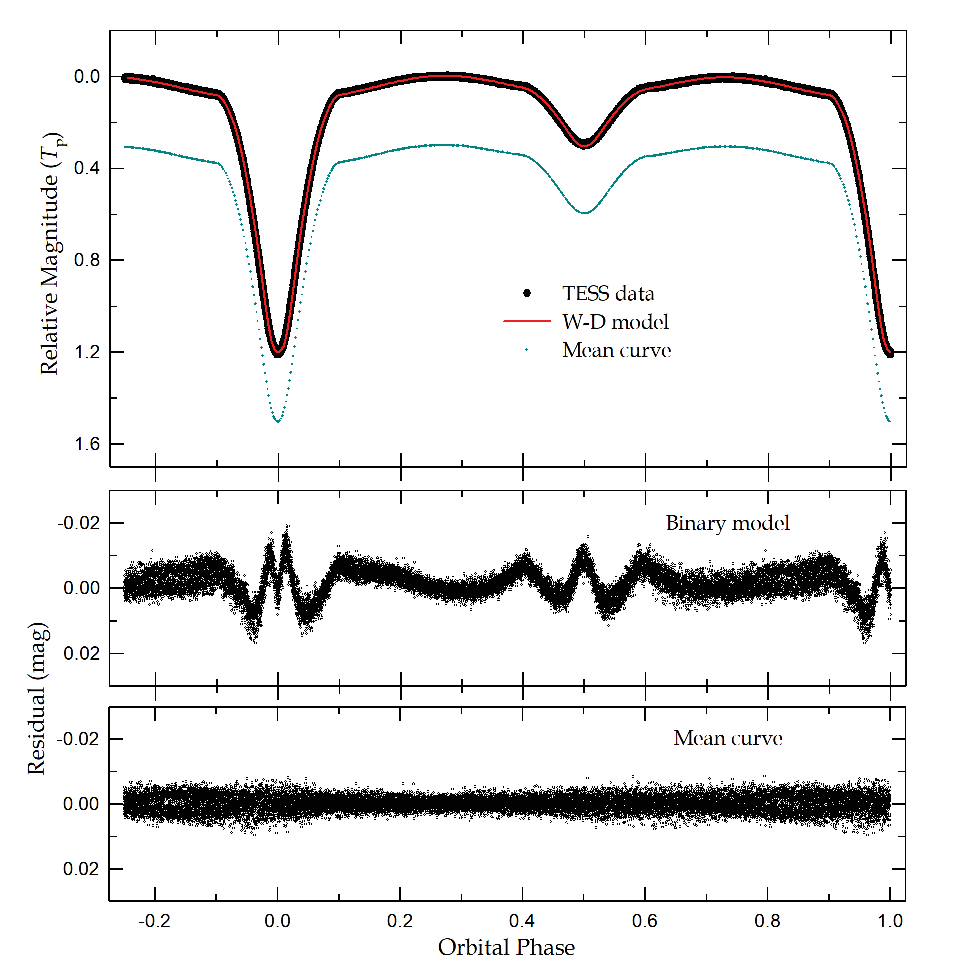}
\end{center}
\caption{Light curve of XZ UMa with a fitted model. In the top panel, the black circles are individual measurements from TESS, 
and the red line represents the synthetic curve obtained from the W-D binary model. The corresponding residuals from this model 
are in the middle panel. The mean curve phase-binned at intervals of 0.002 is plotted as cyan dots in the top panel and displaced 
vertically for clarity. The residuals from the mean light curve are shown in the bottom panel. }
\label{Fig1}
\end{figure}

\begin{figure}
\begin{center}
\includegraphics[]{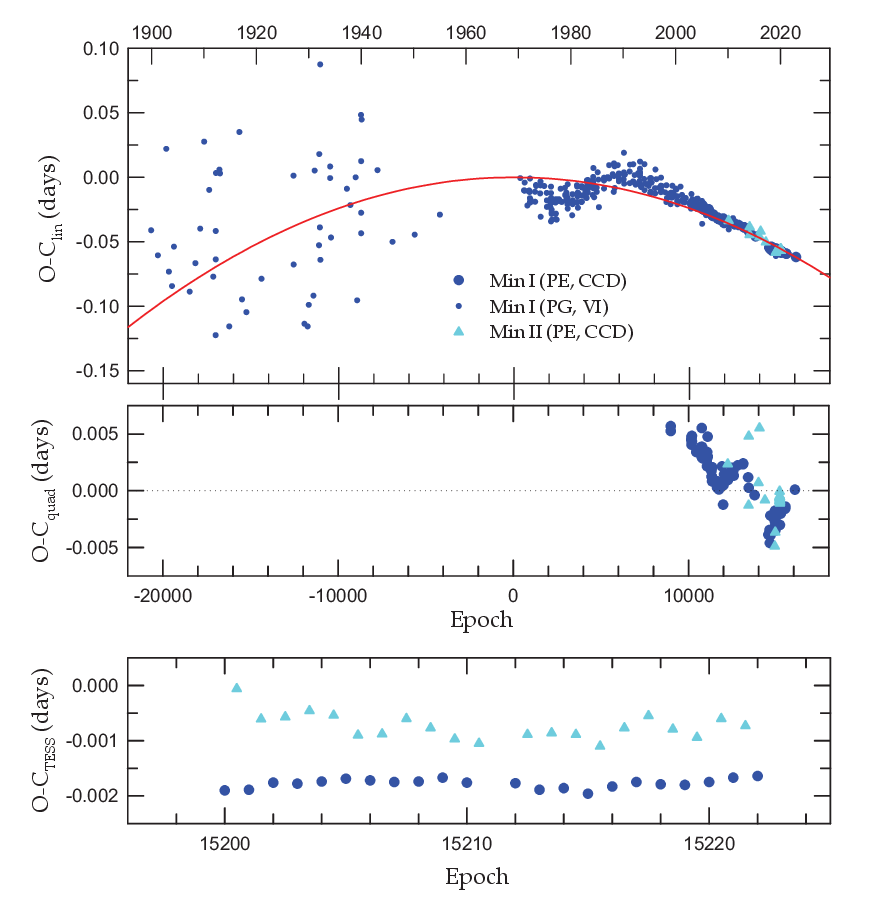}
\end{center}
\caption{Historical $O-C$ diagram of XZ UMa since its discovery, spanning over 120 years of observations. The individual primary minima 
are denoted by blue circles and the secondary by cyan triangles. PE, CCD, PG, and VI stand for photoelectric, CCD, photographic, and 
visual minima, respectively. In the top panel constructed with the linear terms of equation (2), the formal quadratic fit is given 
by the red parabola. After subtracting the parabolic trend, the timing residuals for modern data measured since 2000 are presented in 
the middle panel. The quasi-sinusoidal variations of the $O-C$ values with an amplitude of about 0.004 days are also visible. 
The bottom panel shows only the precise TESS mid-eclipse times obtained in Jan/Feb 2020 (cf. Table 1).} 
\label{Fig2}
\end{figure}

\begin{figure}
\begin{center}
\includegraphics{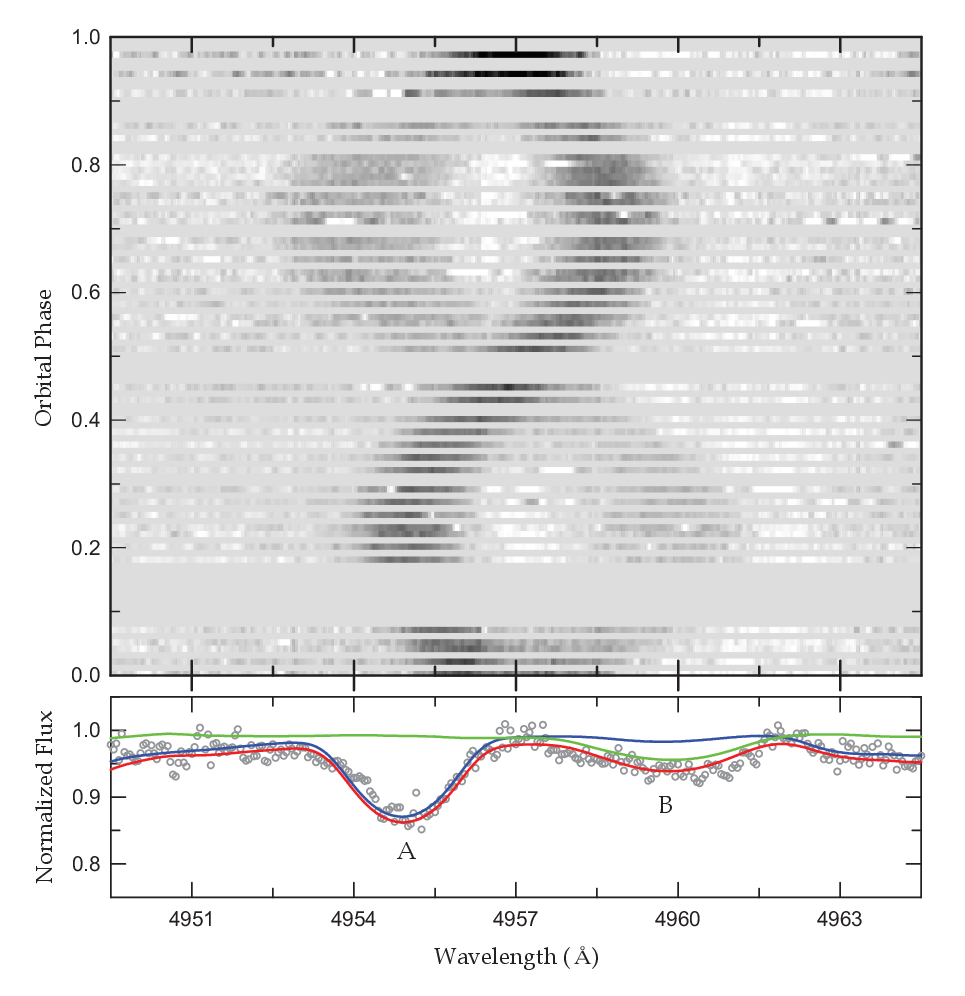}
\end{center}
\caption{Upper panel displays the trailed spectra of XZ UMa in the Fe I $\lambda 4957$ region. In the lower panel, the circle is 
the observed spectrum at an orbital phase of 0.256 (BJD 2,459,554.2000), the blue, green, and red lines represent the synthetic 
spectra of the primary (A; $T_{\rm eff,A}$ = 7940 K, $\log$ $g_{\rm A}$ = 4.28, [M/H] = $-0.15$, $v_{\rm A}$$\sin$$i$ = 80 km s$^{-1}$) 
and secondary (B; $T_{\rm eff,B}$ = 5160 K, $\log$ $g_{\rm B}$ = 3.78, [M/H]$=-0.15$, $v_{\rm B}$$\sin$$i$ = 103 km s$^{-1}$) components, 
and their convolution spectrum, respectively. }
\label{Fig3}
\end{figure}

\begin{figure}
\begin{center}
\includegraphics[]{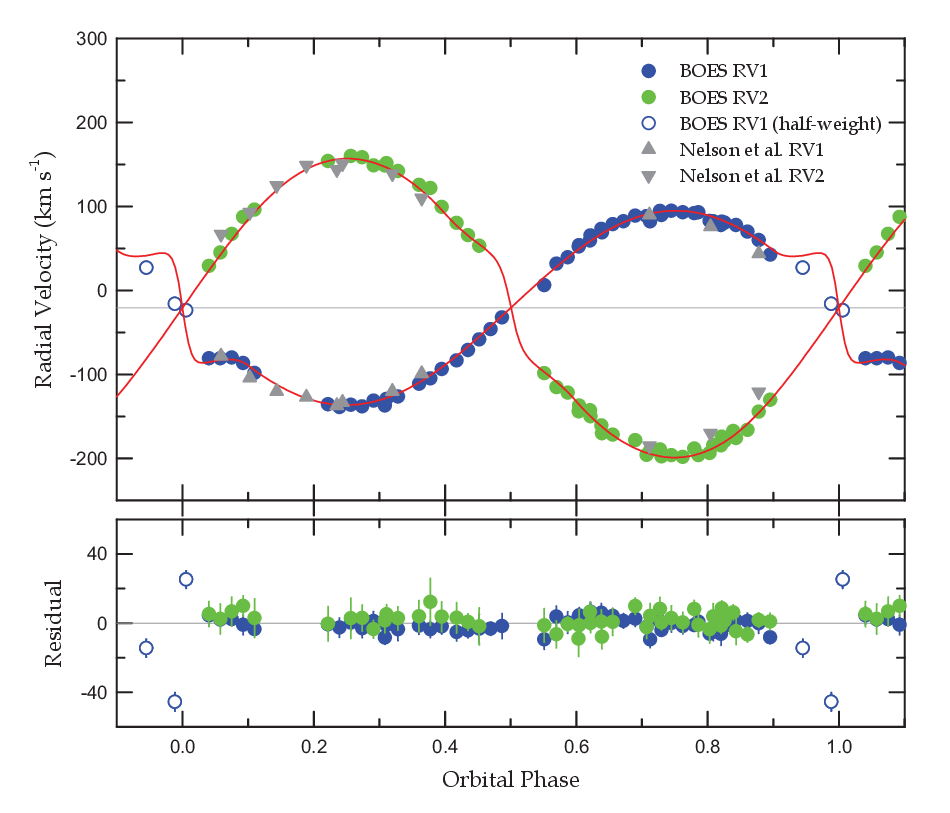}
\end{center}
\caption{RV curves of XZ UMa with fitted models. The blue and green circles are our measurements for the primary and 
secondary stars, respectively, and the red solid curves represent the results from a consistent light and RV curve analysis with 
the W-D code. The open symbols indicate observations given half-weights in the solution. The gray line in the upper panel denotes 
the system velocity of $-$20.74 km s$^{-1}$. The lower panel shows the residuals between observations and models. For comparison, 
the RV measures by Nelson et al. (2006) are displayed as gray triangles in the upper panel. }
\label{Fig4}
\end{figure}

\begin{figure}
\begin{center}
\includegraphics[]{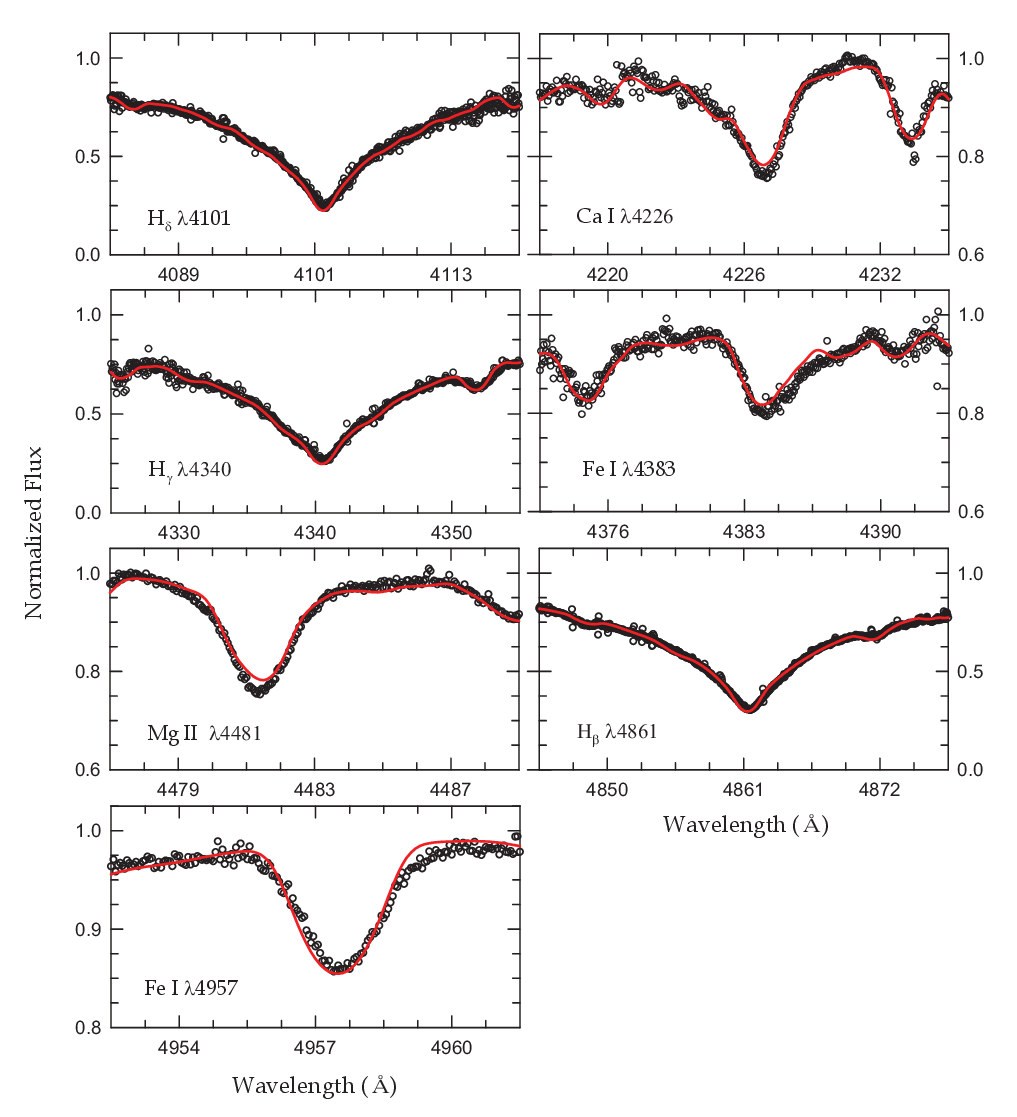}
\end{center}
\caption{Seven spectral regions of the primary star XZ UMa A. The black circles show the disentangled spectrum obtained with 
the FDB\textsc{inary} code. The red lines denote the synthetic spectrum with $T_{\rm eff,A}$ = 7940 K, $\log$ $g_{\rm A}$ = 4.28, 
[M/H] = $-0.15$, $v_{\rm A}$$\sin$$i$ = 80 km s$^{-1}$. } 
\label{Fig5}
\end{figure}

\begin{figure}
\begin{center}
\includegraphics[]{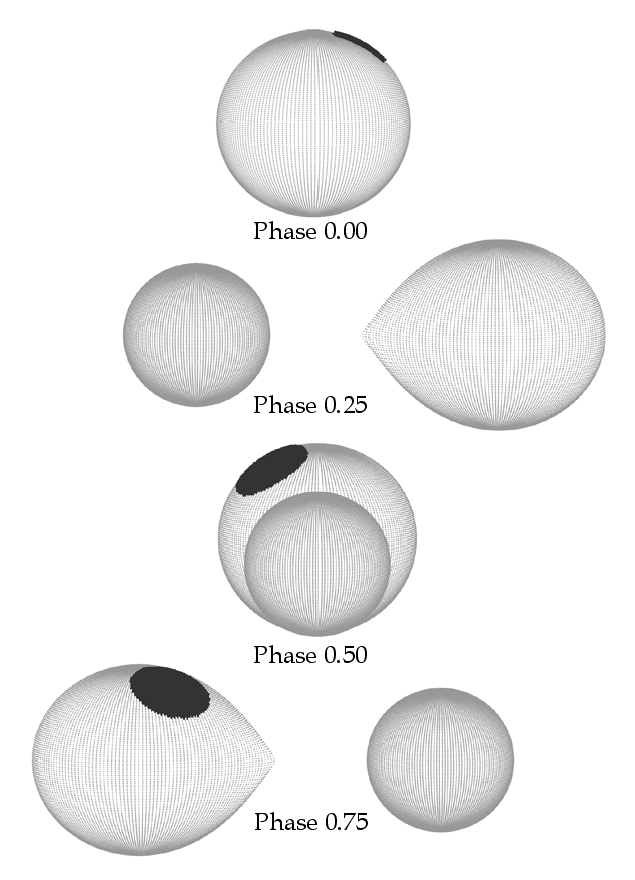}
\end{center}
\caption{Geometrical representations of the Roche lobe surfaces of XZ UMa at four characteristic phases. The smaller of the two stars 
is the primary component, and the black dots indicate a cool spot on its surface. }
\label{Fig6}
\end{figure}

\begin{figure}
\begin{center}
\includegraphics[]{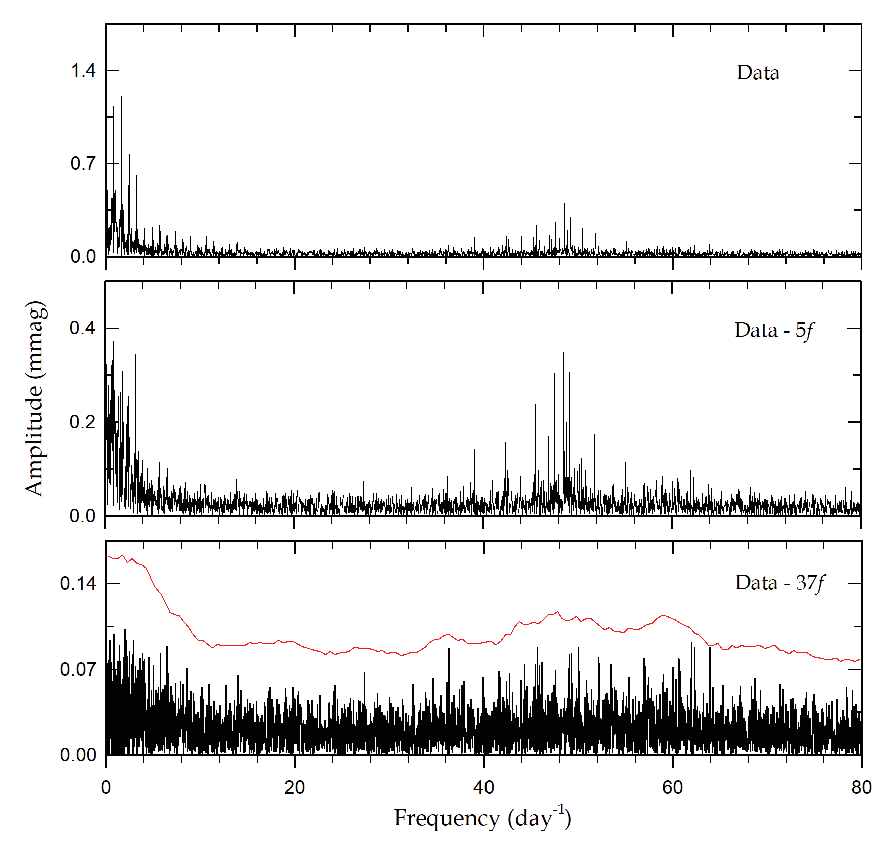}
\end{center}
\caption{PERIOD04 periodograms for the light curve residuals from both the W-D binary model (top panel) and the mean curve (second panel). 
The vertical dotted lines shown in the upper two panels denote a series of orbital multiples up to $85f_{\rm orb} \simeq$ 69.541 day$^{-1}$, 
where $f_{\rm orb}$ = 0.8181281 day$^{-1}$. The third and bottom panels represent the amplitude spectra after pre-whitening the first five 
frequencies and then 32 frequencies for the mean-curve residuals, respectively. The red line in the bottom panel corresponds to four times 
the noise spectrum. }
\label{Fig7}
\end{figure}

\clearpage
\begin{table}
\tbl{TESS Eclipse Timings for XZ UMa. }{%
\begin{tabular}{lccccc}
\hline
BJD              & Error           & Min$\rm ^a$   & BJD              & Error           & Min$\rm ^a$  \\                    
\hline 
2,458,870.62049  & $\pm$0.00002    & I	           & 2,458,885.28825  & $\pm$0.00003    & I            \\ 
2,458,871.23348  & $\pm$0.00013    & II            & 2,458,885.90028  & $\pm$0.00009    & II           \\ 
2,458,871.84280  & $\pm$0.00003    & I             & 2,458,886.51043  & $\pm$0.00002    & I            \\ 
2,458,872.45523  & $\pm$0.00012    & II            & 2,458,887.12262  & $\pm$0.00010    & II           \\ 
2,458,873.06524  & $\pm$0.00003    & I             & 2,458,887.73277  & $\pm$0.00003    & I            \\ 
2,458,873.67758  & $\pm$0.00008    & II            & 2,458,888.34489  & $\pm$0.00010    & II           \\ 
2,458,874.28752  & $\pm$0.00003    & I             & 2,458,888.95497  & $\pm$0.00003    & I            \\ 
2,458,874.89999  & $\pm$0.00010    & II            & 2,458,889.56698  & $\pm$0.00007    & II           \\ 
2,458,875.50986  & $\pm$0.00003    & I             & 2,458,890.17740  & $\pm$0.00002    & I            \\ 
2,458,876.12221  & $\pm$0.00010    & II            & 2,458,890.78961  & $\pm$0.00008    & II           \\ 
2,458,876.73221  & $\pm$0.00003    & I             & 2,458,891.39979  & $\pm$0.00003    & I            \\ 
2,458,877.34415  & $\pm$0.00009    & II            & 2,458,892.01214  & $\pm$0.00010    & II           \\ 
2,458,877.95449  & $\pm$0.00003    & I             & 2,458,892.62205  & $\pm$0.00002    & I            \\ 
2,458,878.56648  & $\pm$0.00009    & II            & 2,458,893.23420  & $\pm$0.00009    & II           \\ 
2,458,879.17676  & $\pm$0.00002    & I             & 2,458,893.84434  & $\pm$0.00002    & I            \\ 
2,458,879.78906  & $\pm$0.00009    & II            & 2,458,894.45635  & $\pm$0.00008    & II           \\
2,458,880.39907  & $\pm$0.00003    & I             & 2,458,895.06669  & $\pm$0.00002    & I            \\
2,458,881.01119  & $\pm$0.00008    & II            & 2,458,895.67899  & $\pm$0.00010    & II           \\
2,458,881.62144  & $\pm$0.00002    & I             & 2,458,896.28908  & $\pm$0.00002    & I            \\
2,458,882.23330  & $\pm$0.00008    & II            & 2,458,896.90117  & $\pm$0.00011    & II           \\
2,458,882.84366  & $\pm$0.00002    & I             & 2,458,897.51141  & $\pm$0.00002    & I            \\
2,458,883.45552  & $\pm$0.00008    & II            &                  &                 &              \\
\hline
\end{tabular}}\label{tab:1}
\begin{tabnote}
\footnotemark[a]Type of eclipse. 
\end{tabnote}
\end{table}

\begin{table}
\tbl{Radial velocities of XZ UMa. }{%
\begin{scriptsize}
\begin{tabular}{lrrrr}
\hline
BJD                    & $V_{\rm A}$        & $\Delta V_{\rm A}$ & $V_{\rm B}$        & $\Delta V_{\rm B}$   \\ 
                       & (km s$^{-1}$)      & (km s$^{-1}$)      & (km s$^{-1}$)      & (km s$^{-1}$)        \\                    
\hline 
2,459,251.1358         & $ -129.2 $         &  3.3               & $   151.6 $        &      5.6             \\
2,459,251.1570         & $ -126.3 $         &  6.5               & $   142.2 $        &      6.7             \\
2,459,251.1963         & $ -111.2 $         &  4.9               & $   125.7 $        &      9.1             \\
2,459,251.2175         & $ -104.7 $         &  4.8               & $   122.0 $        &     13.6             \\
2,459,251.2387         & $  -93.6 $         &  4.3               & $    99.5 $        &      8.6             \\
2,459,251.2663         & $  -83.2 $         &  5.1               & $    80.4 $        &      8.8             \\
2,459,251.2875         & $  -71.0 $         &  4.3               & $    65.9 $        &      4.8             \\
2,459,251.3087         & $  -58.3 $         &  8.4               & $    53.2 $        &     10.7             \\
2,459,251.3299         & $  -46.1 $         &  3.8               & $         $        &                      \\
2,459,251.3511         & $  -32.1 $         &  7.3               & $         $        &                      \\
2,459,252.9389         & $   93.1 $         &  4.5               & $  -196.4 $        &      6.8             \\
2,459,252.9600         & $   82.8 $         &  3.9               & $  -193.9 $        &      8.0             \\
2,459,252.9812         & $   77.9 $         &  6.5               & $  -184.8 $        &      4.4             \\
2,459,253.1333$\rm ^a$ & $   27.1 $         &  5.1               & $         $        &                      \\
2,459,253.1864$\rm ^a$ & $  -15.9 $         &  5.3               & $         $        &                      \\
2,459,253.2076$\rm ^a$ & $  -23.6 $         &  5.1               & $         $        &                      \\
2,459,253.2499         & $  -80.8 $         &  4.9               & $    29.3 $        &      7.3             \\
2,459,253.2711         & $  -80.7 $         &  3.9               & $    45.1 $        &      8.7             \\
2,459,253.2923         & $  -79.8 $         &  3.8               & $    67.3 $        &      8.4             \\
2,459,253.3135         & $  -86.5 $         &  5.7               & $    87.5 $        &      5.9             \\
2,459,253.3347         & $  -98.3 $         &  3.4               & $    96.3 $        &     11.2             \\
2,459,253.9383         & $   52.4 $         &  4.8               & $  -144.0 $        &     10.2             \\
2,459,253.9595         & $   59.5 $         &  4.6               & $  -142.8 $        &      5.6             \\
2,459,253.9807         & $   73.2 $         &  5.4               & $  -160.8 $        &      6.9             \\
2,459,254.0019         & $   79.0 $         &  4.4               & $  -171.6 $        &      7.7             \\
2,459,254.0218         & $   82.3 $         &  4.2               & $         $        &                      \\
2,459,254.0439         & $   89.1 $         &  4.4               & $  -178.3 $        &      4.7             \\
2,459,254.0650         & $   88.5 $         &  6.5               & $  -196.0 $        &      6.5             \\
2,459,254.0899         & $   94.7 $         &  4.0               & $  -189.6 $        &      6.7             \\
2,459,254.1111         & $   94.9 $         &  5.2               & $  -196.4 $        &      7.1             \\
2,459,254.1323         & $   93.0 $         &  4.5               & $  -198.5 $        &      5.9             \\
2,459,254.1535         & $   92.0 $         &  2.6               & $  -188.5 $        &      5.0             \\
2,459,254.1889         & $   82.4 $         &  4.9               & $  -184.4 $        &      7.4             \\
2,459,254.2101         & $   80.7 $         &  3.0               & $  -179.3 $        &      4.0             \\
2,459,254.2313         & $   77.8 $         &  6.6               & $  -175.8 $        &      7.6             \\
2,459,254.2525         & $   70.0 $         &  4.7               & $  -166.1 $        &      4.4             \\
2,459,254.2737         & $   60.1 $         &  5.7               & $  -144.4 $        &      4.2             \\
2,459,254.2948         & $   42.6 $         &  3.6               & $  -130.3 $        &      4.8             \\
2,459,255.2938         & $   82.2 $         &  4.8               & $  -190.8 $        &      7.2             \\
2,459,255.3150         & $   89.9 $         &  3.3               & $  -197.7 $        &      4.2             \\
2,459,303.9887         & $    6.2 $         &  5.7               & $   -98.4 $        &      9.7             \\
2,459,524.3333         & $   81.9 $         &  4.1               & $  -174.2 $        &      4.3             \\
2,459,524.3545         & $        $         &                    & $  -167.4 $        &      4.4             \\
2,459,552.1384         & $   31.9 $         &  6.1               & $  -115.2 $        &      7.8             \\
2,459,552.1596         & $   39.5 $         &  6.2               & $  -121.7 $        &      4.3             \\
2,459,552.1808         & $   54.0 $         &  4.5               & $  -137.0 $        &      9.1             \\
2,459,552.2020         & $   65.6 $         &  5.6               & $  -149.8 $        &      5.6             \\
2,459,552.2231         & $   69.2 $         &  3.8               & $  -170.1 $        &      7.0             \\
2,459,554.1576         & $ -135.5 $         &  3.7               & $   154.2 $        &     10.0             \\
2,459,554.1788         & $ -138.8 $         &  5.6               & $         $        &                      \\
2,459,554.2000         & $ -136.1 $         &  4.5               & $   160.0 $        &     11.7             \\
2,459,554.2212         & $ -138.1 $         &  5.4               & $   158.7 $        &      7.5             \\
2,459,554.2424         & $ -131.3 $         &  5.5               & $   148.9 $        &      5.1             \\
2,459,554.2635         & $ -137.0 $         &  3.8               & $   148.8 $        &      7.1             \\
\hline
\end{tabular}\label{tab:2}
\end{scriptsize}}
\begin{tabnote}
\footnotemark[a]Half-weight given in the binary modeling. 
\end{tabnote}
\end{table}

\begin{table}
\tbl{Light and RV Parameters of XZ UMa. }{%
\begin{tabular}{lcc}
\hline
Parameter                          & Primary                        & Secondary                            \\ 
\hline  
$T_0$ (BJD)                        & \multicolumn{2}{c}{2,458,882.843680$\pm$0.000022}                     \\
$P_{\rm orb}$ (day)                & \multicolumn{2}{c}{1.2223008$\pm$0.0000012}                           \\
$a$ ($R_\odot$)                    & \multicolumn{2}{c}{7.325$\pm$0.043}                                   \\
$\gamma$ (km s$^{-1}$)             & \multicolumn{2}{c}{$-$20.74$\pm$0.69}                                 \\
$K_1$ (km s$^{-1}$)                & \multicolumn{2}{c}{116.0$\pm$1.1}                                     \\
$K_2$ (km s$^{-1}$)                & \multicolumn{2}{c}{186.2$\pm$1.4}                                     \\
$q$                                & \multicolumn{2}{c}{0.6231$\pm$0.0075}                                 \\
$i$ (deg)                          & \multicolumn{2}{c}{85.15$\pm$0.35}                                    \\
$T_{\rm eff}$ (K)                  & 7940$\pm$120$\rm ^a$           & 5162$\pm$53                          \\
$\Omega$                           & 4.811$\pm$0.069                & 3.106                                \\
$F$                                & 1.10$\pm$0.10$\rm ^a$          & 1.0$\rm ^a$                          \\
$A$                                & 1.0$\rm ^a$                    & 0.749$\pm$0.039                      \\
$g$                                & 1.0$\rm ^a$                    & 0.684$\pm$0.043                      \\
$x$, $y$                           & 0.583$\pm$0.080, 0.213$\rm ^a$ & 0.598$\pm$0.067, 0.204$\rm ^a$       \\
$l$/($l_{1}$+$l_{2}$+$l_{3}$)      & 0.608$\pm$0.002                & 0.338                                \\
$l_{3}$$\rm ^b$                    & \multicolumn{2}{c}{0.054$\pm$0.016}                                   \\
$r$ (pole)                         & 0.2378$\pm$0.0042              & 0.3173$\pm$0.0054                    \\
$r$ (point)                        & 0.2451$\pm$0.0047              & 0.4515$\pm$0.0077                    \\
$r$ (side)                         & 0.2411$\pm$0.0044              & 0.3317$\pm$0.0056                    \\
$r$ (back)                         & 0.2439$\pm$0.0046              & 0.3638$\pm$0.0062                    \\
$r$ (volume)$\rm ^c$               & 0.2410$\pm$0.0044              & 0.3390$\pm$0.0058                    \\ [0.5mm]
\multicolumn{3}{l}{Spot Parameters:}                                                                       \\
Colatitude (deg)                   &                                & 36.7$\pm$4.6                         \\        
Longitude (deg)                    &                                & 301.6$\pm$2.2                        \\        
Radius (deg)                       &                                & 23.6$\pm$3.0                         \\        
$T$$\rm _{spot}$/$T$$\rm _{local}$ &                                & 0.936$\pm$0.022                      \\
\hline
\end{tabular}}\label{tab:3}
\begin{tabnote}
\footnotemark[a]Fixed parameter. \\
\footnotemark[b]Value at 0.25 orbital phase. \\
\footnotemark[c]Mean volume radius. 
\end{tabnote}
\end{table}

\begin{table}
\tbl{Absolute Parameters of XZ UMa. }{%
\begin{tabular}{lccccc}
\hline
Parameter                     & \multicolumn{2}{c}{Nelson et al. (2006)}    && \multicolumn{2}{c}{This Paper}               \\ [0.5mm] \cline{2-3} \cline{5-6}   
                              & Primary           & Secondary               && Primary           & Secondary                \\    
\hline 
$M$ ($M_\odot$)               & 1.92$\pm$0.09     & 1.20$\pm$0.05           && 2.177$\pm$0.040   & 1.356$\pm$0.026          \\
$R$ ($R_\odot$)               & 1.70$\pm$0.03     & 2.38$\pm$0.04           && 1.764$\pm$0.034   & 2.481$\pm$0.045          \\
$\log$ $g$ (cgs)              & 4.26$\pm$0.02     & 3.76$\pm$0.02           && 4.283$\pm$0.018   & 3.781$\pm$0.017          \\
$\rho$ ($\rho_\odot$)         & \,                & \,                      && 0.397$\pm$0.024   & 0.089$\pm$0.005          \\
$v_{\rm sync}$ (km s$^{-1}$)  & \,                & \,                      && 73.0$\pm$1.4      & 102.7$\pm$1.9            \\
$v$$\sin$$i$ (km s$^{-1}$)    & \,                & \,                      && 80$\pm$7          & \,                       \\
$T_{\rm eff}$ (K)             & 7766$\pm$240      & 5346$\pm$5              && 7940$\pm$120      & 5162$\pm$53              \\
$L$ ($L_\odot$)               & 9.5$\pm$0.1       & 4.2$\pm$0.1             && 11.08$\pm$0.79    & 3.92$\pm$0.21            \\
$M_{\rm bol}$ (mag)           & \,                & \,                      && 2.119$\pm$0.078   & 3.248$\pm$0.059          \\
BC (mag)                      & \,                & \,                      && 0.021$\pm$0.001   & $-$0.241$\pm$0.019       \\
$M_{\rm V}$ (mag)             & \,                & \,                      && 2.098$\pm$0.078   & 3.489$\pm$0.063          \\
Distance (pc)                 & \multicolumn{2}{c}{504$\pm$26}              && \multicolumn{2}{c}{489$\pm$20}               \\
\hline
\end{tabular}}\label{tab:4}
\end{table}

\begin{table}
\tbl{Results of the multiple frequency analysis for XZ UMa$\rm ^a$. }{%
\begin{tabular}{lrccrc}
\hline
             & Frequency              & Amplitude           & Phase           & S/N$\rm ^b$            & Remark                  \\ [-2.0ex]
             & (day$^{-1}$)           & (mmag)              & (rad)           &                        &                         \\
\hline
$f_{1}$      &  0.8575$\pm$0.0003     & 1.533$\pm$0.066     & 3.06$\pm$0.13   & 40.05                  & $1f_{\rm orb}$          \\
$f_{2}$      &  1.6675$\pm$0.0006     & 0.877$\pm$0.068     & 4.05$\pm$0.23   & 22.12                  & $2f_{\rm orb}$          \\
$f_{3}$      &  0.1024$\pm$0.0006     & 0.812$\pm$0.066     & 5.72$\pm$0.24   & 20.94                  &                         \\ 
$f_{4}$      &  0.0366$\pm$0.0006     & 0.784$\pm$0.066     & 5.89$\pm$0.25   & 20.27                  & data timespan           \\ 
$f_{5}$      &  1.0404$\pm$0.0015     & 0.317$\pm$0.066     & 0.74$\pm$0.61   &  8.28                  &                         \\
$f_{6}$      & 48.5399$\pm$0.0009     & 0.373$\pm$0.045     & 3.91$\pm$0.36   & 14.10                  &                         \\ 
$f_{7}$      &  0.8831$\pm$0.0013     & 0.357$\pm$0.066     & 3.66$\pm$0.54   &  9.29                  & $1f_{\rm rot}$          \\
$f_{8}$      &  3.2437$\pm$0.0011     & 0.442$\pm$0.064     & 0.78$\pm$0.42   & 11.86                  & $4f_{\rm orb}$          \\
$f_{9}$      &  0.7058$\pm$0.0010     & 0.473$\pm$0.065     & 4.66$\pm$0.40   & 12.42                  &                         \\ 
$f_{10}$     &  2.4245$\pm$0.0020     & 0.233$\pm$0.064     & 4.92$\pm$0.81   &  6.22                  & $3f_{\rm orb}$          \\
$f_{11}$     &  1.8358$\pm$0.0018     & 0.264$\pm$0.066     & 4.84$\pm$0.74   &  6.82                  & $2f_{\rm rot}$          \\
$f_{12}$     & 49.1670$\pm$0.0010     & 0.339$\pm$0.045     & 1.23$\pm$0.39   & 12.81                  &                         \\ 
$f_{13}$     &  0.1408$\pm$0.0015     & 0.320$\pm$0.066     & 3.82$\pm$0.61   &  8.27                  &                         \\ 
$f_{14}$     &  0.9983$\pm$0.0016     & 0.293$\pm$0.065     & 3.59$\pm$0.65   &  7.69                  &                         \\ 
$f_{15}$     &  0.3346$\pm$0.0017     & 0.279$\pm$0.066     & 2.66$\pm$0.69   &  7.21                  &                         \\ 
$f_{16}$     &  1.4463$\pm$0.0018     & 0.265$\pm$0.066     & 1.14$\pm$0.73   &  6.85                  &                         \\ 
$f_{17}$     &  1.1501$\pm$0.0019     & 0.259$\pm$0.066     & 4.43$\pm$0.74   &  6.73                  &                         \\ 
$f_{18}$     & 45.5467$\pm$0.0015     & 0.220$\pm$0.044     & 0.25$\pm$0.59   &  8.56                  &                         \\ 
$f_{19}$     &  0.9142$\pm$0.0014     & 0.343$\pm$0.066     & 5.81$\pm$0.56   &  8.93                  &                         \\ 
$f_{20}$     &  0.7515$\pm$0.0015     & 0.322$\pm$0.065     & 0.12$\pm$0.59   &  8.45                  &                         \\ 
$f_{21}$     &  0.6601$\pm$0.0021     & 0.227$\pm$0.065     & 4.02$\pm$0.84   &  5.99                  &                         \\ 
$f_{22}$     &  2.2453$\pm$0.0018     & 0.263$\pm$0.064     & 4.89$\pm$0.72   &  7.01                  &                         \\ 
$f_{23}$     & 50.4579$\pm$0.0018     & 0.178$\pm$0.044     & 2.79$\pm$0.73   &  6.85                  &                         \\ 
$f_{24}$     & 43.9925$\pm$0.0023     & 0.142$\pm$0.044     & 1.43$\pm$0.91   &  5.53                  &                         \\ 
$f_{25}$     &  0.2286$\pm$0.0028     & 0.170$\pm$0.066     & 4.97$\pm$1.14   &  4.39                  &                         \\ 
$f_{26}$     & 45.8959$\pm$0.0020     & 0.159$\pm$0.044     & 2.65$\pm$0.81   &  6.17                  & $f_{12}-4f_{\rm orb}$   \\ 
$f_{27}$     & 48.0242$\pm$0.0021     & 0.159$\pm$0.046     & 4.10$\pm$0.85   &  5.90                  & $f_{12}-f_{17}$         \\ 
$f_{28}$     &  0.5705$\pm$0.0026     & 0.184$\pm$0.065     & 5.73$\pm$1.04   &  4.83                  &                         \\ 
$f_{29}$     & 39.0886$\pm$0.0019     & 0.139$\pm$0.037     & 0.87$\pm$0.78   &  6.41                  &                         \\ 
$f_{30}$     & 42.6029$\pm$0.0022     & 0.132$\pm$0.039     & 4.15$\pm$0.87   &  5.75                  & $f_{26}-4f_{\rm orb}$   \\ 
$f_{31}$     & 51.8128$\pm$0.0027     & 0.119$\pm$0.044     & 6.10$\pm$1.08   &  4.65                  & $f_{6}+4f_{\rm orb}$    \\ 
$f_{32}$     &  4.8801$\pm$0.0027     & 0.157$\pm$0.059     & 3.74$\pm$1.09   &  4.61                  & $6f_{\rm orb}$          \\
\hline 
\end{tabular}}\label{tab:5}
\begin{tabnote}
\footnotemark[a]Frequencies are listed in order of detection. \\
\footnotemark[b]Calculated in a range of 5 day$^{-1}$ around each frequency.
\end{tabnote}
\end{table}

\end{document}